\documentclass[preprint,NumberedRefs]{JASA}

\begin{document}

\title[preprint submitted to JASA]{Polyvinylidene fluoride transducer shape optimization for the characterization of anisotropic materials}
\author{Diego A. Cowes}
\email{diegocowes@cnea.gov.ar}
\affiliation{Departamento ICES, Comisión Nacional de Energía Atómica, Villa Maipú, B1650, Argentina}
\affiliation{Instituto Sabato, Universidad Nacional de San Martín, Villa Maipú, B1650, Argentina}

\author{Juan I. Mieza}
\affiliation{Hidrógeno en materiales, Comisión Nacional de Energía Atómica, Villa Maipú, B1650, Argentina}
\affiliation{Instituto Sabato, Universidad Nacional de San Martín, Villa Maipú, B1650, Argentina}

\author{Martín P. Gómez}
\affiliation{Departamento ICES, Comisión Nacional de Energía Atómica, Villa Maipú, B1650, Argentina}
\affiliation{Instituto Sabato, Universidad Nacional de San Martín, Villa Maipú, B1650, Argentina}
\affiliation{Departamento de Mecánica, Universidad Tecnológica Nacional, Campana, B2804, Argentina}

\preprint{Diego A. Cowes, JASA}	

\date{\today}

\begin{abstract}
In the context of the ultrasonic determination of mechanical properties, it is common to use oblique incident waves to characterize fluid-immersed anisotropic samples. The lateral displacement of the ultrasonic field owing to leaky guided wave phenomena poses a challenge for data inversion because beam spreading is rarely well represented by plane-wave models. In this study, a finite beam model based on the angular spectrum method was developed to estimate the influence of the transducer shape and position on the transmitted signals. Additionally, anisotropic solids were considered so that the beam skewing effect was contemplated. A small-emitter large-receiver configuration was chosen, and the ideal shape and position of the receiving transducer were obtained through a meta-heuristic optimization approach with the goal of achieving a measurement system that sufficiently resembles plane-wave propagation. A polyvinylidene fluoride receiver was fabricated based on the findings and tested in three cases: a single-crystal silicon wafer, a lightly anisotropic stainless-steel plate, and a highly anisotropic composite plate. Good agreement was found between the measurements and the plane-wave model.  
\end{abstract}


\maketitle


\section{\label{sec:1} Introduction}
Ultrasonic methods are typically used to determine the elastic constants of generally anisotropic elastic solids, owing to the coupling between mechanical properties and wave propagation phenomena. Frequently, ultrasonic waves are transmitted through a material of interest while varying the angle between the transducer and sample. These experiments are commonly referred to as ultrasonic goniometry. The acquired pulse time-of-flight \citep{Rokhlin1992, Aristgui1997, Martens2019}, critical angle \citep{Rokhlin1989}, transmitted spectra \citep{Castaings2000, Hosten2008}, dispersion curves \citep{Fei2003, Holland2004, Bochud2018},  or waveforms \citep{Leymarie2002} can then be used in an inversion procedure to determine the stiffness tensor. 

\begin{figure}[t]
\includegraphics[width=0.95\reprintcolumnwidth]{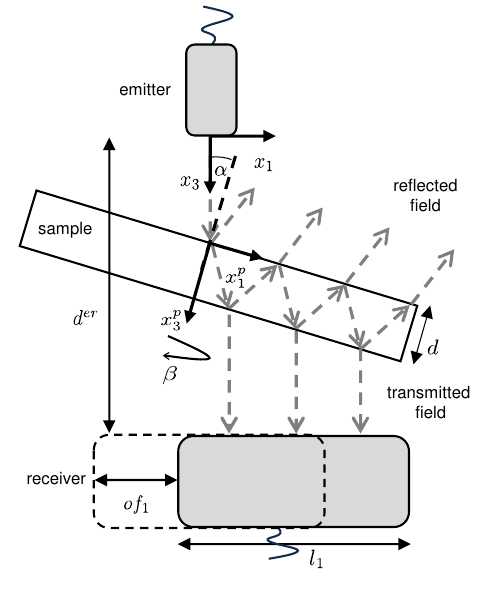}
\caption{\label{fig:FIG1}{Diagram of an ultrasonic goniometry experiment. Grey dashed arrows show the leaky guided wave phenomena.}}
\end{figure}

A challenge of ultrasonic goniometry is that oblique incident waves travel back and forth in the material, which acts as a waveguide, resulting in the lateral displacement of successive echoes \citep{Georgiades2022}, as depicted in Fig.\ref{fig:FIG1}. Plane wave models account for this, but  are rarely well-adjusted to the experiments because the finite dimensions of the receiver transducer fail to acquire all the displaced energy.
Finite beam models can account for this phenomenon but require the double integration of all the wavevector components owing to beam diffraction \citep{Lobkis1999, Potel2005}, which significantly increases the computational complexity and time. Because fitting the model to the measured field is performed by an iterative approach that requires numerous evaluations, a measurement setup that sufficiently resembles plane wave propagation would avoid the need to use a finite beam model, thereby greatly reducing the number of operations.

An early workaround called the synthetic plane wave technique \citep{Hosten1993, Jocker2007} consisted of simulating an infinite receiver transducer by laterally scanning and summing the received signals.
Later investigations used larger transducers to achieve better agreement between the measured transmission coefficients and plane-wave theory without the need to laterally scan the field \citep{Castaings2000, Bouzidi2006, Cawley1997}.
Furthermore, later works \citep{Adamowski2008, Kumar2006, Puthillath2010} delved into the design of large-aperture receivers fabricated from polyvinylidene ﬂuoride (PVDF),  which was shown to be a suitable choice for this application.

PVDF is a piezoelectric polymer that has several advantages over ceramic elements. Specifically, its internal losses produce a damped oscillatory response, resulting in short pulses with excellent temporal resolution, which is useful for transit time determination, and a broadband spectrum that is well-suited for transmission and reflection coefficient measurements. Additionally, its low acoustic impedance is similar to that of water, which eliminates the need for impedance-matching layers \citep{Brown2000}. 
However, owing to its low dielectric constant, high dielectric losses, and low electromechanical coupling coefficient, the capacitance of PVDF dominates the acoustic load \citep{Swartz1980}. The transducer capacitance, which is a function of the area, indicates that the transducer response is highly dependent on the transducer area \citep{Sherar1989}. Thus, although a larger transducer surface better approximates a plane wave model, it deviates from the optimal capacitance, thereby reducing the sensitivity and bandwidth.

This work aims to develop a PVDF transducer with a constrained area whose response resembles plane wave propagation by optimizing the transducer shape and position.
To achieve this, a model based on the decomposition of plane waves is developed in Section \ref{sec:2}. This model simulates ultrasonic fields in the temporal and frequency domains, and the signals acquired by finite transducers in such fields. The model analysis in the context of ultrasonic goniometry is presented in Section \ref{sec:3}. The choice of using a small emitter and large receiver over other configurations is justified in Subsection \ref{subsec:3:2}. The reduction in error between the plane wave propagation and the finite-beam model when increasing the transducer length and lateral offset is shown in Subsections \ref{subsec:3:4} and \ref{subsec:3:5}, respectively. The shape optimization process and its implementation in graphic processing units (GPU) are described in Section \ref{sec:4}. Finally, the constructed transducer is evaluated in Section \ref{subsec:5:2} using real materials of known and unknown properties.

\section{\label{sec:2} Theoretical Basis}
\subsection{\label{subsec:2:1} Angular spectrum}
For a non-viscous fluid that does not resist shear deformation, the potential associated with the particle velocity can be expressed in the form of a monochromatic plane wave as
\begin{equation}\label{eq:1}
    \phi=\Phi \exp(i(k_1x_1+k_2x_2+k_3x_3-\omega t)),
\end{equation}
where $\Phi$ is the potential amplitude, $k_i$ is the wavevector component in the $x_i$ direction,  $\omega$ is the angular frequency, and $t$ is time. The monochromatic finite field at the face of an immersed emitting transducer can be decomposed as the summation of plane waves, suppressing the temporal term $\exp(-i\omega t)$, as
\begin{equation}\label{eq:2}
    \phi(x_1,x_2,0)=\frac{1}{4\pi^2}\iint\Phi^e(k_1,k_2)\exp(i(k_1x_1+k_2x_2))\text{d}k_1\text{d}k_2,
\end{equation}
where $\Phi^e$ is the potential amplitude associated with the emitting transducer that determines the emitter directivity. If the transducer acts as a piston and the velocity potential is known at its face, the directivity is obtained as
\begin{equation}\label{eq:3}
    \Phi^e(k_1,k_2)=\iint\phi(x_1,x_2,0)\exp(-i(k_1x_1+k_2x_2))\text{d}x_1\text{d}x_2.
\end{equation}
This decomposition allows the calculation of the field at a parallel plane ($x_3 \neq 0)$ through the propagating factor $H=\exp(ik_3x_3)$ as
\begin{multline}\label{eq:4}
    \phi(x_1,x_2,x_3)=\frac{1}{4\pi^2}\iint\Phi^e(k_1,k_2)\exp(i(k_1x_1+k_2x_2))
    \\ \cdot \exp(ik_3x_3)\text{d}k_1\text{d}k_2,
\end{multline}
where, for isotropic media, the $k_3$ wavevector component is obtained from
\begin{equation}\label{eq:5}
k_3 =
    \begin{cases}
        \sqrt{k^2-k_1^2-k_2^2} & \text{if }  k^2\ge k_1^2+k_2^2 \\
        i\sqrt{k_1^2+k_2^2-k^2} & \text{if } k^2 < k_1^2+k_2^2,
    \end{cases}
\end{equation}
where $k=\omega / c$ is the wavenumber and $c$ is the speed of sound in the fluid.
When real, it describes propagating waves, and when imaginary, it describes inhomogeneous waves that decay exponentially with distance. Therefore, inhomogeneous waves are only present in the vicinity of the transducer face; thus, imaginary wavevector components can be omitted in the calculations \citep{Schmerr2007}.
This procedure is called the Angular Spectrum Method 
(ASM), and one of its advantages is that it can be implemented via the Fast Fourier Transform algorithm (FFT), which is highly optimized and consequently fast, thus Eq. \ref{eq:3} can be rewritten as
\begin{equation}\label{eq:6}
     \Phi^e(k_1,k_2)=FFT\{\phi(x_1,x_2,0)\},
\end{equation}
and Eq. \ref{eq:4} can be rewritten as
\begin{eqnarray}
  \phi(x_1,x_2,x_3)&=&iFFT\{FFT\{\phi(x_1,x_2,0)\}H(k_1,k_2,x_3)\} \nonumber \\
   & = & iFFT\{\Phi^e(k_1,k_2)H(k_1,k_2,x_3)\}. 
\end{eqnarray}
The assumption of a piston transducer, i.e. the same particle velocity potential at every surface element in the transducer face, is not a requirement of the angular spectrum and a different profile can be implemented in Eq. \ref{eq:2}. More sophisticated  finite element models \citep{Aanes2016, Mosland2023}, may be used to estimate such a profile, but for this work the piston assumption is considered sufficient.

\subsection{\label{subsec:2:2}Signal reception}
Through the Kino-Auld reciprocity relations\citep{Auld1979, Kino1978, Atalar1988}, the signal produced by a receiving transducer insonified by a finite beam can be expressed as
\begin{multline}\label{eq:8}
   S^{fb}(d^{er},\omega)=\frac{\rho_f\omega}{8\pi^2P(\omega)}\iint k_3 \Phi^e(k_1,k_2)\exp(ik_3d^{er})\\
   \cdot \Phi^r(-k_1,-k_2) \text{d}k_1 \text{d}k_2,
\end{multline}
where the super index $fb$ refers to the finite beam, $d^{er}$ is the emitter-receiver distance, $\rho_f$ is the fluid density,  $P$ is the incident power at the transducer terminals, and $\Phi^r(-k_1,-k_2)$ is the directivity of the receiving transducer obtained through Eq. \ref{eq:3}.

A second advantage of the ASM is that the scattering coefficients at fluid-solid interfaces are usually obtained for plane waves. Consequently, a finite beam decomposed into plane waves can be propagated through an interface by multiplying each wave with the scattering coefficient. If a rotated anisotropic solid plate is placed between the transducers, the transmission coefficient relates the incident and transmitted potentials amplitudes as $\Phi^I(k_1,k_2)T(k_1',k_2')=\Phi^T(k_1,k_2)$
where the plate rotation is considered at the wavevectors $(k_1'(k_1,k_2),k_2'(k_1,k_2))$  by the passive rotation described in Section \ref{subsec:2:4}. Inhomogeneous waves also appear at fluid–solid interfaces, but these are not discarded because they are a critical part of $T$. The transmitted field at the receiver plane is obtained as
\begin{multline}\label{eq:9a}
    \phi(x_1,x_2,d^{er})=\frac{1}{4\pi^2}\iint\Phi^e(k_1,k_2)\exp(i(k_1x_1+k_2x_2))
    \\ \cdot T(k_1',k_2')\exp(ik_3d^{er})\text{d}k_1\text{d}k_2,
\end{multline}
and the signal generated by the receiver in this field is
\begin{multline}\label{eq:9}
   S^{fb}(d^{er},\omega)=\frac{\rho_f\omega}{8\pi^2P(\omega)}\iint k_3 \Phi^e(k_1,k_2)T(k_1',k_2')\\ 
   \cdot \exp(ik_3d^{er})\Phi^r(-k_1,-k_2) \text{d}k_1 \text{d}k_2.
\end{multline}
Eqs. \ref{eq:9} and \ref{eq:8} represent the finite beam model (FBM) with and without a solid plate respectively.

Conversely, for a transducer approaching an infinite area, its directivity approaches a delta function such that only the normal component $k_1=k_2=0$ remains, and the plane wave propagation is recovered. Then the double integral reduces to
\begin{equation}\label{eq:10}
   S^{pw}(d^{er},\omega)=\frac{\rho_f\omega}{8\pi^2P(\omega)} k_3(0,0) T'(0,0)\exp(ik_3(0,0)d^{er}),
\end{equation}
where the super index $pw$ refers to the plane wave and $k_3(0,0)=k$ and $T'(0,0)=T(k_1'(0,0),k_2'(0,0))$. Considering $\Gamma(\omega)$ as the system's impulse response acquired in the absence of the plate, the plane wave model can be abbreviated as
\begin{equation}\label{eq:11}
     S^{pw}(d^{er},\omega)=\Gamma(\omega)T(\omega).
\end{equation}
Then, the difference between models is named as finite beam error (FBE) and is described as 
\begin{equation}\label{eq:12}
        FBE=(|S^{pw}(d^{er},\omega)|-|S^{fb}(d^{er},\omega)|)^2.
\end{equation}
At sufficiently low errors, Eq. \ref{eq:11} can be used instead of Eq. \ref{eq:9} avoiding the burden of the double integration, which is the main objective of this investigation.

\subsection{\label{subsec:2:3}Pulsed fields}
Typically, ultrasonic fields are bounded both in space and in time, thus the analysis is expanded from monochromatic to wideband behavior by recovering the temporal term $\exp(-i\omega t)$ with an additional Fourier transform from frequency to time for the particle velocity as
\begin{equation}\label{eq:13}
        \phi(x_1,x_2,x_3,t)=\int \phi(x_1,x_2,x_3,\omega)\exp(-i\omega t)\text{d}\omega,
\end{equation}
and for the received signal as
\begin{equation}\label{eq:14}
        s(d^{er},t)=\int S(d^{er},\omega)\exp(-i\omega t)\text{d}\omega.
\end{equation}
\subsection{\label{subsec:2:4}Coordinate system and passive rotation}
In this study, a coordinate system located at the emitter transducer face is adopted, as depicted in \ref{fig:FIG1}. Because of this, each rotation of the plate requires the passive rotation of the wavevectors for the calculation of the transmission coefficient $T(k_1',k_2')$ described as
\begin{equation}\label{eq:15}
    \begin{bmatrix}       
    k_1' \\k_2'\\k_3'
    \end{bmatrix}=
         \begin{bmatrix}
    \cos(\beta)&-\sin(\beta)&0\\
    \sin(\beta)&\cos(\beta)&0\\
    0&0&1\\
    \end{bmatrix}
    \begin{bmatrix}
    \cos(\alpha)&0&\sin(\alpha)\\
    0&1&0\\
    -\sin(\alpha)&0&\cos(\alpha)\\
    \end{bmatrix}
    \begin{bmatrix}
        k_1 \\k_2\\k_3
    \end{bmatrix}.
\end{equation}
This represents  a rotation about an axis perpendicular to the plate normal, $x_2^p$ (polar rotation by $\alpha$), followed by a rotation around the plate normal $x_3^{p}$ (azimuthal rotation by $\beta$). This strategy enables the use of a compact receiver because beam displacement is predominately observed in the positive $x_1$ direction, regardless of the rotation angles. 
$\alpha$ and $\beta$ are chosen to represent the spherical coordinates instead of the more commonly used $\theta$ and $\phi$ letters, because they  usually describe the plane wave angle of incidence, which, for finite beams, may differ from the angles between the transducer and sample normals.

\begin{figure}[t]
    \centering
    \includegraphics[width=0.97\reprintcolumnwidth]{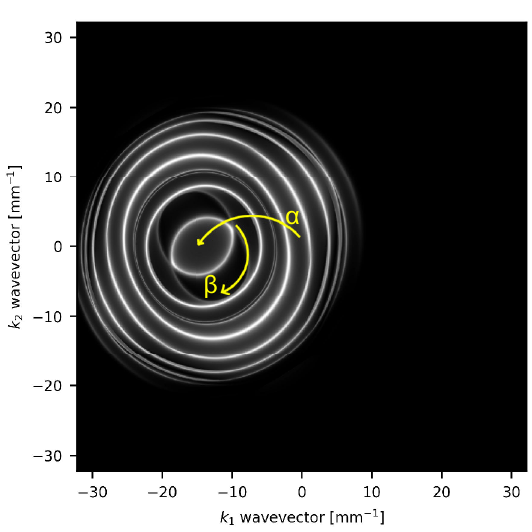}
    \caption{\label{fig:FIG2}Transmission coefficient of an anisotropic steel plate rotated at $\alpha = 20^{\circ}$ and $\beta = 60 ^{\circ}$ for a frequency of $10\ \text{MHz}$. Arrows show the displacement of a point after the rotation.}  
\end{figure}

A reference system located at the plate surface would require the rotation of the ultrasonic field, producing a non-uniform sampling step and negating the use of the standard FFT algorithm \citep{Potel2005}.
However, the drawback of the chosen coordinate system is that for each rotation, the wavevector array changes such that the transmission coefficient needs to be recomputed. Nevertheless, this was preferred for the purpose of this study.  The absolute value of the transmission coefficient was calculated for a slightly anisotropic SAE 304 steel plate with a thickness of $1\ \text{mm}$ whose stiffness tensor is included in \ref{appendix:1}, for rotations of $\alpha = 20^{\circ}$ and $\beta = 60 ^{\circ}$ and a frequency of $10\ \text{MHz}$. The results are shown in Fig. \ref{fig:FIG2}. 

Additionally, the presence of the plate disrupts the travel distance in water, such that the transmission coefficient also includes a phase change $\exp(i\varphi)=\exp(i(k_1 d \sin(\alpha)-k_3d\cos(\alpha)))$, which can be derived from a rotation analogous to that of Eq. \ref{eq:15}. The calculation of the transmission coefficient for anisotropic plates is covered presented in \citep{Nayfeh1988-bn, Rokhlin2002, Kiefer2019}.

\subsection{\label{subsec:2:5}Analytical directivity}
While the FFT depicted in Eq. \ref{eq:6} is applicable for any planar transducer shape, Eq. \ref{eq:3} can be analytically solved for a circular transducer as
\begin{equation}\label{eq:16}
     \Phi(k_1,k_2)=2\pi a^2\phi_0\frac{J_1\Big(a\sqrt{k_1^2+k_2^2}\Big)}{a\sqrt{k_1^2+k_2^2}},
\end{equation}
and for a rectangular transducer as
\begin{equation}\label{eq:17}
    \Phi(k_1,k_2)=l_1l_2\phi_0\frac{\sin(\frac{k_1l_1}{2})\sin(\frac{k_2l_2}{2})}{(\frac{k_1l_1}{2})(\frac{k_2l_2}{2})},
\end{equation}
where $J_1$ is the Bessel function of order 1; $\phi_0$ is a known velocity potential; and $a$, $l_1$,$l_2$ are the transducer radius, length, and width, respectively.
These Eqs. are used in Section \ref{sec:3} when studying the field of transducers with known shapes, whereas the FFT of Eq. \ref{eq:3} is used in section \ref{sec:4} during the transducer shape optimization process where Eq. \ref{eq:3} has no analytical solution.

\section{\label{sec:3}Finite beam model analysis}
\subsection{\label{subsec:3:1}Transducer directivity functions}
\begin{figure}[t]
    \centering
    \includegraphics[width=0.97\reprintcolumnwidth]{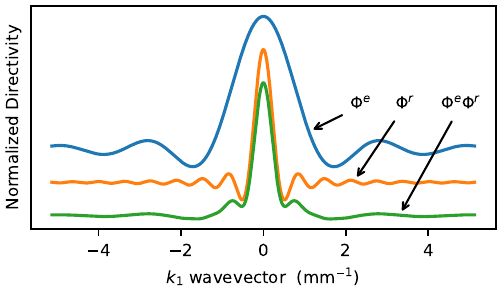}
    \caption{\label{fig:FIG3}Directivity functions of two circular transducers at $10\ \text{MHz}$ in water. $\Phi^e$ corresponds to a radius of 3 mm, $\Phi^r$ corresponds to a radius of 10 mm. $\Phi^e\Phi^rr$ is the directivity product. Each function has a different added offset for improved visualization.(Color online).}
    
\end{figure}

By reviewing Eq. \ref{eq:9} it can be concluded that the FBM approximates the PWM only when the product of the directivity functions $\Phi^e\Phi^r$ approaches a Dirac delta function. This way, the double integral reduces to the plane wave transmission coefficient at the $\alpha$ angle where a single value of $k_1$ and $k_2$ is different from zero.
Eqs. \ref{eq:16} and \ref{eq:17} are sinc-like functions that become narrower when the radius $a$ or the lengths $l_1$ or $l_2$ increase. Thus, increasing the transducer area makes the FBM approach a PWM. However, the key parameter is the directivity product so that a broad directivity function multiplied by a narrow directivity function still produces a narrow product. Fig. \ref{fig:FIG3} displays Eq. \ref{eq:16} calculated at $k_2=0$ for two circular piston transducers for a frequency of $10\ \text{MHz}$ radiating into water ($c=1490\ \text{m/s}$), where the emitter directivity $\Phi^e$ corresponds to a radius of $3\ \text{mm}$ and the receiver directivity $\Phi^r$ corresponds to a radius of $10\ \text{mm}$. The directivities product $\Phi^e\Phi^r$ resembles to the directivity of the bigger transducer.

\begin{figure}[t]
    \centering
    \includegraphics[width=0.97\reprintcolumnwidth]{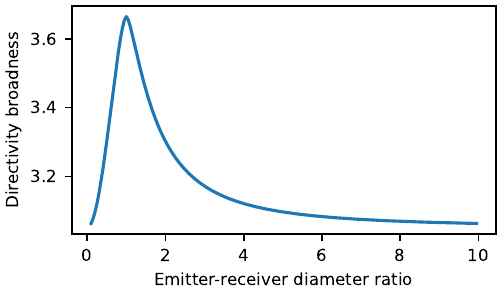}
    \caption{\label{fig:FIG4}Directivity product broadness as a function of emitter-receiver radius ratio and for a fixed total surface.}
    
\end{figure}

Using the area under the curve as a proxy for directivity broadness, Fig. \ref{fig:FIG4} shows the integral of the directivity product against the emitter-receiver radius ratio for a fixed sum of transducer area. The calculations are made for circular pistons at $10\ \text{MHz}$ with a total area of $1000\ \text{mm}^2$ radiating into water. It can be seen that the broader directivity corresponds to the case where both transducers have the same dimensions (ratio=1), suggesting that for a fixed total area, it is advantageous to use transducers of dissimilar dimensions. Moreover, starting from a set of dissimilar transducers, it is more cost-effective to increase the size of the bigger transducer rather than that of the smaller one.
\subsection{\label{subsec:3:2}Field Visualization}
\begin{figure}[t]
    \centering
    \includegraphics[width=0.97\reprintcolumnwidth]{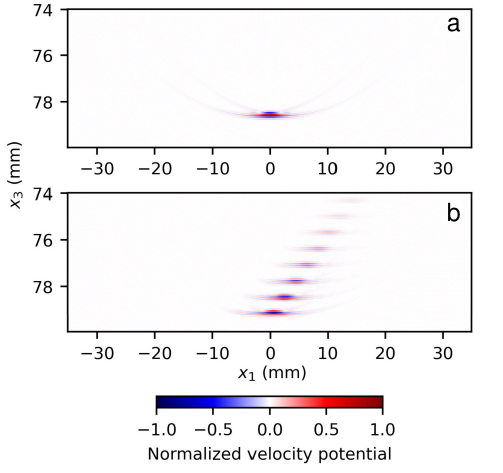}
    \caption{\label{fig:FIG5}Velocity potential field simulation in water at the $x_1x_3$ plane of a pulsed ultrasonic field generated by a 3 mm radius emitter with a central frequency of 10 MHz. Top (a) and bottom (b) figures correspond to the case without and with the presence of a steel plate, respectively (Color online).}
\end{figure}

Eqs. \ref{eq:9a} and \ref{eq:4} can be used to visualize the velocity potential field at a given frequency before the reception process with and without the solid plate, respectively.  Evaluating Eqs. \ref{eq:4} and \ref{eq:9a} through,  the Fourier transform of Eq. \ref{eq:13} can be used to represent the pulsed field at a given time. The fields, generated by a 3 mm radius piston emitter radiating into water, are computed in the $x_1x_3$ reception plane, which is located at $x_3=80\ \text{mm}$.  The transmission coefficient corresponds to the steel plate as calculated in Section \ref{subsec:2:4}. The temporal signal corresponds to a Gaussian pulse with a center frequency of $10\ \text{MHz}$. Fig. \ref{fig:FIG5} shows the lateral field expansion which corresponds to the successive echoes generated by the water-solid interfaces of the immersed plate. As expected, each echo is displaced to the right and delayed by the time it takes to travel back and forth in the material. While in this case the echoes can be visually separated, for different $\alpha$ rotations the echoes can superimpose with each other, which is prevailing in guided wave propagation.

\begin{figure}[t]
    \centering
    \includegraphics[width=0.97\reprintcolumnwidth]{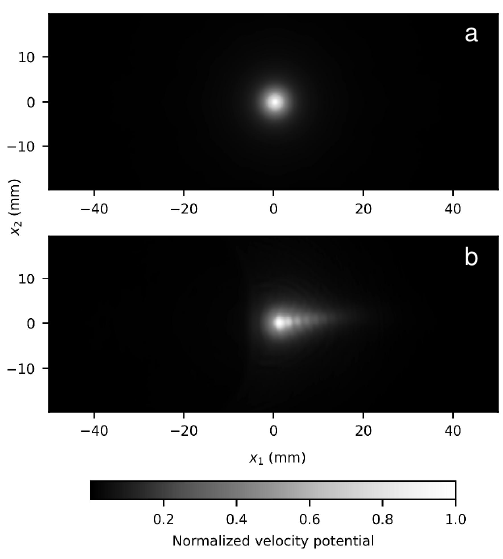}
    \caption{\label{fig:FIG6}Velocity potential field in the $x_1x_2$ plane at $x_3$= 80 mm, for a 3 mm radius emitter, as a summation of all the frequencies of interest. Top (a) and bottom (b) figures correspond to the case without and with the presence of a steel plate, respectively.}    
\end{figure}

\begin{figure}[t]
    \centering
    \includegraphics[width=0.97\reprintcolumnwidth]{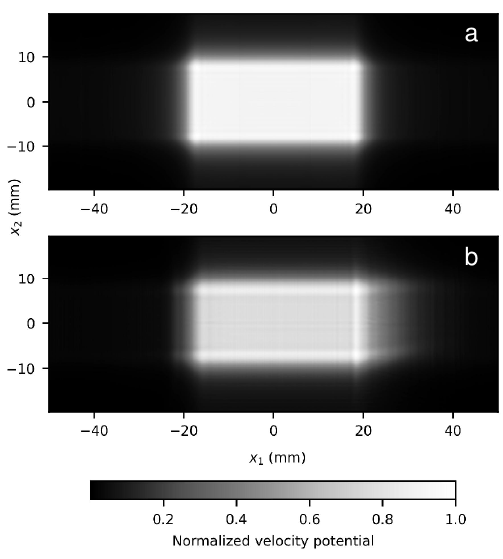}
    \caption{\label{fig:FIG7}Velocity potential field in the $x_1x_2$ plane at $x_3$= 80 mm, for rectangular wide emitter, as a summation of all the frequencies of interest. Top (a) and bottom (b) figures correspond to the case without and with the presence of a steel plate respectively}
\end{figure}

Alternatively, if Eqs. \ref{eq:4} and \ref{eq:9a} are not transformed into the time domain, they can be used for the visualization of the energy impinging the $x_1x_2$ reception plane ($x_3=80\ \text{mm})$. These equations are solved and summed for all the frequencies of interest. 
The conditions are the same as the former simulation, but Fig. \ref{fig:FIG6} is calculated for a circular emitter ($a$ = 3 mm) while Fig. \ref{fig:FIG7} shows the field of a rectangular emitter ($l_1=20\ \text{mm}$ and $l_2= 40\ \text{mm}$).   The top (a) and bottom (b) figures show the field at the reception plane without and with the presence of the solid plate, respectively. It can be seen that the field is expanded to the right ($x_1$) following the scheme shown in Fig. \ref{fig:FIG1}.  A small deviation in the positive $x_2$ direction is observed, which is present due to a beam skewing effect described in Section \ref{subsec:3:6}.


\subsection{\label{subsec:3:3}Emitter-receiver configuration}
Revisiting Eq. \ref{eq:9}  it can be concluded that the emitter and receiver are reciprocal, and theoretically their functions can be interchanged without changing the measured response.  The physical idea behind this can be thought of in two ways: a large emitter produces a flat and wide wavefront that approximates a plane wave, and a small receiver only senses this field at a point in space, resembling the role of a hydrophone; the opposite case is a point emitter which produces a spherical wavefront, while a wide receiver is selective only to the components of the wavefront perpendicular to its surface, effectively filtering out oblique incidence energy. Both configurations, while not producing the same ultrasonic field, generate the same received ultrasonic signal, which ideally resembles that of plane wave propagation. 

Figs. \ref{fig:FIG6} and \ref{fig:FIG7} show that the field corresponding to the bigger emitter produces a wider distribution of energy. Regarding the fact that the plate sample has to be held by some kind of fixture, a wider field could potentially interact with it, producing unwanted reflections or changes in boundary conditions. In other words, although the transducers can be theoretically interchanged, it is preferred to choose the narrower field to diminish the probability of undesired interactions.
Because of this, in this work, the transducer configuration consists of a small ($a$ = 3 mm) circular emitter and a wide PVDF receiver whose shape will be determined in Section \ref{sec:4}. 

\subsection{\label{subsec:3:4}Receiver length}
\begin{figure}[t]
    \centering
    \includegraphics[width=0.97\reprintcolumnwidth]{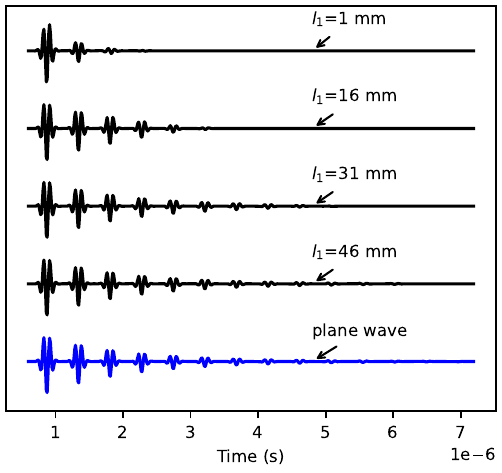}
    \caption{\label{fig:FIG8}Simulation of transmitted signals produced by a finite length emitter and acquired by a rectangular receiver of varying lengths $l_1$.}
\end{figure}

After the field reaches the reception plane, Auld-Kino relations are used to simulate the signal generated at the receiver terminals, which is described by Eqs. \ref{eq:8} and \ref{eq:9}. From Figs. \ref{fig:FIG6} and \ref{fig:FIG7} it can be seen that the field displacement is primarily in the $x_1$ direction, therefore the calculations were performed for a fixed $x_2$ receiver length ($l_2$ = 20 mm) while increasing the $x_1$ receiver length $l_1$. The remaining parameters are the same as in Section \ref{subsec:3:2}.  In accordance with the scheme in Fig. \ref{fig:FIG1}, the signals in Fig. \ref{fig:FIG8} show that the wider the receiver, the more echoes are captured and therefore the FBE decreases.

\begin{figure}[t]
    \centering
    \includegraphics[width=0.97\reprintcolumnwidth]{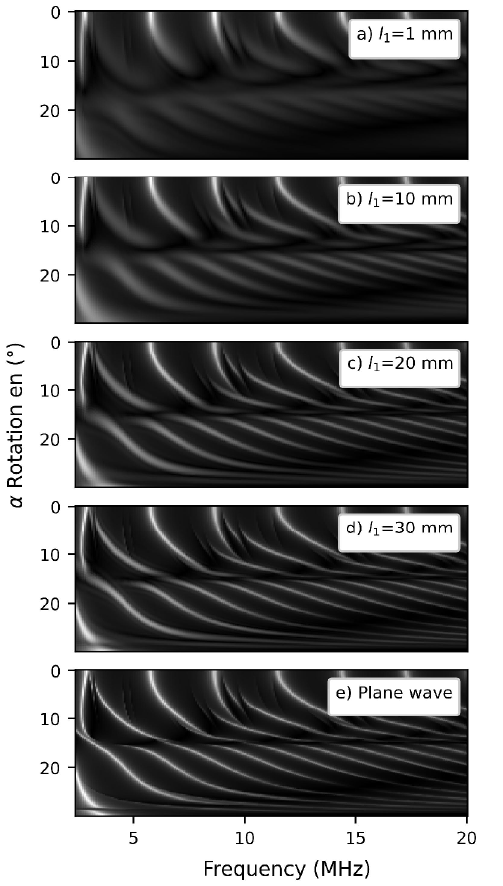}
    \caption{\label{fig:FIG9}Simulation of transmitted spectra produced by a finite length emitter and acquired by a rectangular receiver of different lengths $l_1$, while varying $\alpha$ rotation angle }
\end{figure}

The periodicity of these echoes produces a comb filtering effect, that can be observed in the frequency domain. Because of this, transmitted spectra as a function of rotation angle are generally used to study guided wave and dispersion phenomena.
 The transmitted spectra were simulated using Eq. \ref{eq:9} repeating the above conditions but varying the $\alpha$ rotation angle between 0° and 30°. Again, as shown in Fig. \ref{fig:FIG9} the bigger the receiver $x_1$ length ($l_1$), the lower the FBE. 

\subsection{\label{subsec:3:5}Lateral offset}
\begin{figure}[t]
    \centering
    \includegraphics[width=0.97\reprintcolumnwidth]{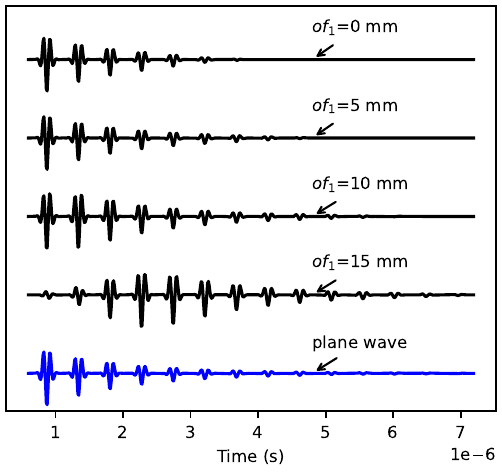}
    \caption{\label{fig:FIG10}Simulation of transmitted signals produced by a finite length emitter and acquired by a rectangular receiver of varying position $of_1$.}
\end{figure}

Although, up to this point, it has become apparent that the use of a larger receiver better resembles the PWM, its position has not yet been addressed. 
Mirroring Section \ref{subsec:3:4}, Eq. \ref{eq:9} is used to study the signal received by a square transducer ($l_1=l_2=20\ \text{mm}$). Here, the length is fixed but the receiver $x_1$-lateral offset is increased from 0 to 15 mm. The results are shown in Fig. \ref{fig:FIG10}. The signals resemble the findings of Jocker \citep{Jocker2007} where the transmitted field is scanned by laterally displacing a conventional transducer. As the offset increases, more echoes are received up to a point in which the amplitude of the early echoes starts to diminish and eventually disappear. This suggests that there is an optimum transducer displacement which reduces the FBE.

\subsection{\label{subsec:3:6}Beam skewing}
\begin{figure}[t]
    \centering
    \includegraphics[width=0.97\reprintcolumnwidth]{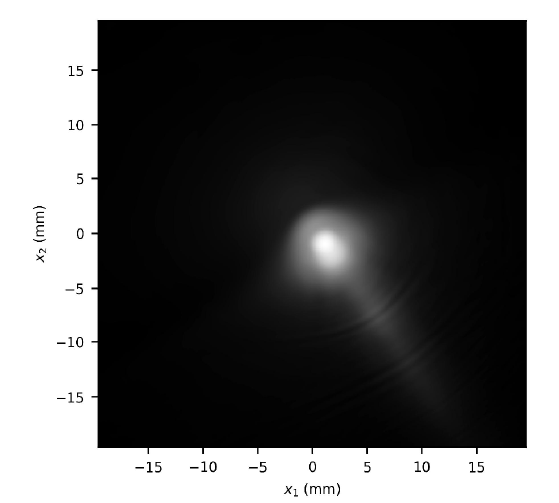}
    \caption{\label{fig:FIG11}Velocity potential field transmitted through a CFRP plate in the $x_1x_2$ plane at $x_3$= 80 mm, for a 3 mm radius emitter, as a summation of all the frequencies of interest. Rotation angles ($\alpha = 20^{\circ}$ and $\beta = 60 ^{\circ}$) are chosen to highlight the beam-skewing effect.}    
\end{figure}

Wave propagation on isotropic materials is characterized by the fact that the wave energy direction is coincident with the wavefront normal. This no longer holds on off-axis planes in anisotropic materials which gives rise to the beam skewing phenomenon, and results in the parting between phase and group velocity \citep{Fromme2018}. 
Beam skewing angles in highly anisotropic materials, such as unidirectional fiber-reinforced composites (CFRP), can be as high as 50°. This is shown in Fig. \ref{fig:FIG11} by repeating the procedure of Section \ref{subsec:3:2} but changing the transmission coefficient to that of a unidirectional CFRP for rotations of $\alpha = 20^{\circ}$ and $\beta = 60 ^{\circ}$. The stiffness tensor is included in Appendix \ref{appendix:1}.

Because the plane of incidence is not one of symmetry,  the energy strongly deviates from the $x_1$ direction towards the $x_2$ direction, which means that for this particular rotation a receiver elongated in the $x_1$ direction may be ill-suited. Nevertheless, as the skewing angle greatly depends on plate rotation, the best shape for all orientations is yet to be determined. Because of this, a receiver shape optimization procedure was conducted for the transmission coefficient of lightly anisotropic stainless steel, a strongly anisotropic unidirectional CFRP and a single-crystal silicon wafer.

\section{\label{sec:4}Optimization procedure}
The qualitative nature of beam spreading in ultrasonic goniometry experiments has been covered in Section \ref{sec:3}, which showed that increasing receiver length ($l_1$) and offset ($of_1$) decreases the FBE. This section aims to quantitatively optimize the receiver shape and position while constraining its area. 
To do this, an optimization procedure was implemented in which the receiver shape is varied and the error between FBM and  PWM is minimized. This was executed by the Differential Evolution algorithm, a stochastic parallel direct search method introduced by Storn and Price based on an evolutionary process, which has been shown to be effective for global optimization \citep{Storn1997} and it was used through the Python pymoo library \citep{pymoo} implementation. A typical optimization consists of 40 individuals which evolve for a total of 500 generations.

The optimization procedure can be summarized in the following steps:
\begin{enumerate}
  \item A population of 40 individuals is randomly generated.
  \item Each individual consists of 10 vertices, which define a closed polygon. The polygon is fitted with a spline to smooth the edges. The enclosed area represents a receiver surface.
  \item The receiver surface is FFT transformed to obtain the directivity function, $\Phi^r(-k_1,-k_2)$ as in Eq. \ref{eq:3}.
  \item The finite beam model is evaluated for each individual, at a grid of 9 $\alpha$ by 13 $\beta$ rotations, 256 $k_1$ by 256 $k_2$ wavevectors, and 64 $\omega$ frequencies. Because the emitter directivity and transmission coefficient are independent of the individual, they are already pre-computed and retrieved from memory for the calculation of Eq. \ref{eq:9} which mainly consists of the multiplication and summation of these arrays with the individual's directivity function.  
\item  The error between models (FBE) is computed according to \ref{eq:12}.
\item  The error is used by the Differential Evolution algorithm to generate the subsequent population. It also imposes the constraint that the polygon area remains under $700\ \text{mm}^2$.
\item  At the end of the optimization, the individual with the least error is reported.
\end{enumerate}

Because of the high dimensionality of the problem ($\alpha(9), \beta(13), k_1(256), k_2(256),\omega(64)$), coupled to the fact that each iteration is very similar to the next, the inversion was implemented for graphics processing units (GPU) architecture, leading to a great reduction in computation time. For this purpose, the CuPy library was used \citep{cupy-learningsys2017}. 

The optimization procedure was carried out for three materials of interest: a lightly anisotropic SAE 304 stainless steel, a single-crystal silicon, and a unidirectional CFRP, whose reference stiffness matrices were found in literature and are included in Appendix \ref{appendix:1}. All materials were taken to be 1 mm in thickness. Due to the elastic symmetry of the materials, the $\beta$ angle ranged from -90° to 90°, while the $\alpha$ angle ranged from 0° to 45° because above this angle little energy is transmitted. The transmission coefficient $T(k_1',k_2')$ was pre-calculated as a matrix of $\alpha$ by $\beta$ by $k_1$ by $k_y$ by $\omega$ and retrieved for each iteration. The emitter directivity $\Phi^e(k_1,k_2)$ was calculated for a circular piston ($a$ =3 mm).

\section{\label{sec:5}Results and discussion}
\subsection{\label{subsec:5:1}Optimization results}
\begin{figure}[t]
    \centering
    \includegraphics[width=0.97\reprintcolumnwidth]{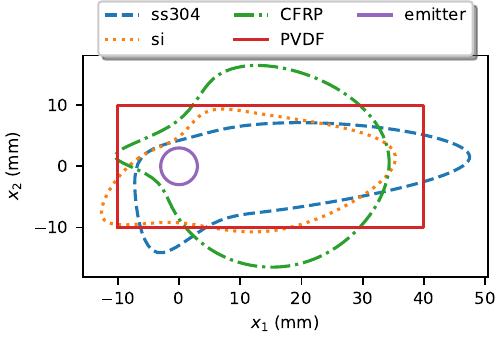}
    \caption{\label{fig:FIG12}Results of the optimization procedure. “ss304”, “si”, and “CFRP” are the shapes optimized for steel, silicon, and composite, respectively. The emitter is included as a reference, and the “PVDF” rectangular shape represents the fabricated receiver.(Color online).}
\end{figure}

The optimized shapes are shown in Fig. \ref{fig:FIG12}, where the emitter is displayed for reference.   The results obtained for stainless steel are consistent with the simulations of sec \ref{subsec:3:2}  in that the shape is elongated in the positive $x_1$ direction and its center is also displaced in the $x_1$ direction. The fact that the shape is not symmetric to the $x_2$ direction may be related to the skewing effect. 

The shape optimized for silicon is similar to that of steel, in that it is elongated in the $x_1$ direction. Nevertheless, the $l_1$ length is smaller, suggesting that the successive echoes decay more rapidly. This may be explained by the lower density of silicon which, being closer to that of water, means that the energy is leaked more efficiently into the fluid.

 The shape optimized for CFRP presents some notorious differences with respect to the former: first, the shape is wider in the $x_2$ direction near the center with a higher radial symmetry, because of this the $x_1$ length is smaller. These differences may be explained by two phenomena. Again, the lower density of the CFRP means that the energy is leaked into the fluid more efficiently than for steel and silicon due to the smaller discrepancy in acoustic impedance, meaning that the amplitude is attenuated more quickly and fewer echoes exist. At the same time, the beam skewing effect due to the pronounced anisotropy makes the energy more dispersed in space so that the radial symmetry of the receiver has to be higher to acquire such a field. 
 
\subsection{\label{subsec:5:2}Transducer construction and evaluation}
Because of the discrepancy between optimized shapes, a rectangular shape of $l_1=50\ \text{mm}$ and $l_2= 20\ \text{mm}$ was selected to coarsely approximate the results, as displayed in Fig. \ref{fig:FIG12}. While not ideal, this shape represents a viable compromise from which the wide PVDF receiver transducer was fabricated.

\begin{figure}[t]
    \centering
    \includegraphics[width=0.7\reprintcolumnwidth]{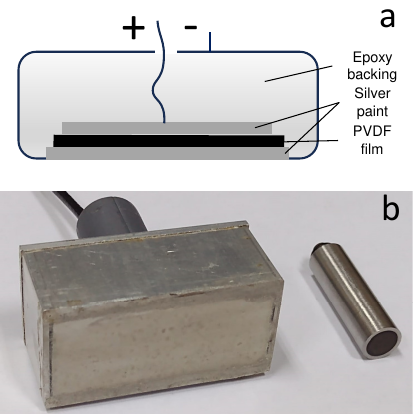}
    \caption{\label{fig:FIG13}(a) Fabrication details of the PVDF receiver.(b) Fabricated PVDF receiver (left) alongside a commercial ceramic emitter (right).}
\end{figure}

The piezoelectric receiver was constructed by using a $28 \ \mu \text{m}$ thick PVDF film with silver paint electrodes. The sensor was placed in an aluminum casing, where epoxy resin was poured and used as a backing material. The front electrode was extended to the casing through the use of silver paint to provide electric contact and radio frequency shielding. The rear electrode was connected to a cable with a small drop of silver paint. Finally, a thin coat of epoxy resin was applied over the front face to provide corrosion protection. A scheme of the construction is shown in Fig. \ref{fig:FIG13}(a), and the constructed transducer is displayed in Fig. \ref{fig:FIG13}(b) alongside a commercial ceramic emitter.

\begin{figure*}[t]
    \centering
    \fig{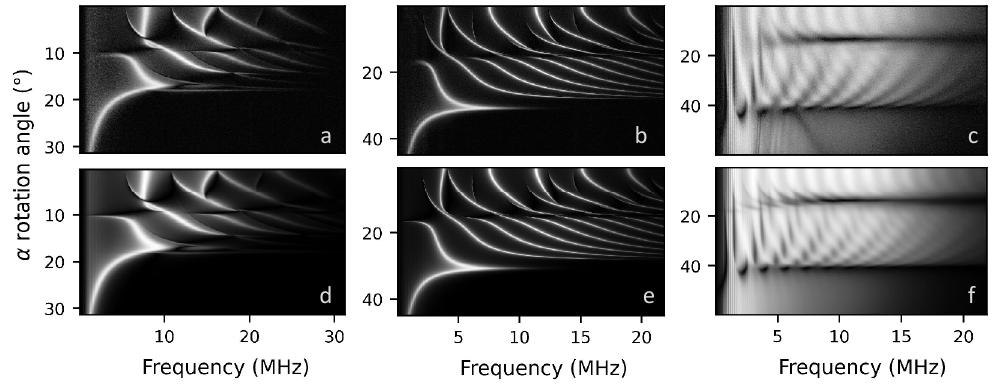}{1.97\reprintcolumnwidth}{}
    \caption{PVDF receiver test. Comparison between measured (a, b, c) and plane wave simulated (d, e, f) transmission spectra for a silicon wafer (a, d), stainless steel (b, e) and CFRP (c, f).}
    \label{fig:FIG14}
\end{figure*}

To evaluate the performance of the developed receiving transducer, actual measurements of three real materials were made. The signal spectrum was acquired by scanning the polar angle $\alpha$ at a fixed $\beta$ angle. A 3 mm radius emitter was used and the transmitter-receiver distance was set at 80 mm. 
The first sample is a 0.5 mm thick single crystal silicon wafer used for micro-electromechanical systems (MEMS) fabrication, whose mechanical properties are well known and are included in Appendix \ref{appendix:1}. The results of the measurements are compared to a plane wave simulation in Fig. \ref{fig:FIG14} (a, d). 

The second sample is a lightly anisotropic 304 stainless steel plate with a thickness of 0.8 mm, and the third sample is a highly anisotropic unidirectional CFRP plate with a thickness of 0.9 mm. The properties of these Appendix \ref{appendix:2}. This procedure will be described in future publications. The comparison between measurement and PWM for a single polar scan is shown in Figs. \ref{fig:FIG14} (b, e) and \ref{fig:FIG14} (c, f) for the steel and composite plates, respectively.

The measurement of the silicon sample, being a material of known properties, serves as proof that the system behavior sufficiently resembles plane wave propagation. The other materials, of unknown properties, show that the plane wave model can be fitted to the measurement, suggesting that the measurement setup is useful for ultrasonic stiffness inversion. Although the construction of arbitrary transducers can be achieved with PVDF films, finding an optimum shape may be unfeasible if material properties are not known.  For these cases, increasing $x_1$ length and $x_1$ offset are useful guidelines for the construction of a receiving transducer. The dimensions presented here serve well for materials of 1 mm thickness or lower. Thicker samples may require longer receivers due to the geometrical nature of wave refraction.

The shape optimization procedure may be useful when material properties are known and their variation under the action of a degradation agent, such as irradiation, is to be determined, for example in the context of fitness for service and non-destructive evaluations \citep{Singh2018, Straalsund1973}.

For the present case, the low discrepancy observed means that plane wave propagation can be assumed and thus, the computation time can be greatly decreased. Because of this, the finite beam calculations carried out in this work with a grid of  $2^8$ by $2^8$  wavevectors, provided plane wave propagation, are reduced to a single one.

\section{\label{sec:6}Conclusion}
This work has shown a theoretical model to simulate ultrasonic fields transmitted through fluid-immersed anisotropic plates. Both spatial and temporal boundedness were considered such that this model is able to simulate fields in the frequency and the temporal domains, as well as the signals generated by a receiver positioned in such fields.  The chosen coordinate system  located at the emitter face permits the calculation of the field for arbitrarily shaped planar transducers by a simple FFT, avoiding the need to rotate the ultrasonic field. 

The developed model was used to show that infinite transducers approach plane wave propagation; that, when accounting for total surface, a set of dissimilar area transducers is preferred over that of equal area; and that due to sample fixing limitations, a small emitter-large receiver configuration is desirable over the opposite one. In this configuration, it was shown that receiver $x_1$-length and $x_1$-offset are critical and can be manipulated to simulate plane wave propagation.

A meta-heuristic optimization procedure was implemented to find the receiving transducer shape that best approximates plane wave propagation while constraining the total area. Two shapes were optimized for different materials, one for a lightly anisotropic stainless steel plate and one for a highly anisotropic unidirectional CFRP plate. The obtained shapes are elongated for the metal and quasi-circular for the composite. This difference may be due to the difference in density, which determines how rapidly energy leaks into the fluid, and to the beam skewing effect predominant in the composite sample.

Lastly, a PVDF receiver was manufactured taking into account the present findings. Actual measurements were then compared to the plane wave model for real materials of known and unknown properties. Excellent agreement was found, indicating that plane wave propagation can be assumed, hence significantly reducing computational complexity and time in field inversions.

\begin{acknowledgments}
The authors wish to thank Diego Perez for the provision of the silicon wafer, Augusto Bonelli Toro for the provision of the composite samples, and the IAEA for the provision of an ultrasonic pulser-receiver in the context of the Technical Cooperation Project called “Building General Capacity in Nuclear Science and Applications in National Strategic Areas”. 
\end{acknowledgments}
\section*{Author Declarations}
The authors have no conflicts to disclose. 
\section*{Data Availability}
The data that support the findings of this study are available from the corresponding author upon reasonable request. 

\appendix
\renewcommand{\arraystretch}{0.7}
\section{Material properties from literature}\label{appendix:1}
Stainless steel 304 grade stiffness matrix in Voigt notation (units=GPa), $\rho=7.81\ \mathrm{g/cm^3}$ \citep{Lan2018-rus}.
\begin{equation}\label{eq:ss}
    \begin{bmatrix}
    259.0&107.0&112.8 &0&0&0\\
    \,    &269.1&101.6 &0&0&0\\
    \,    &\,    & 261.3 &0&0&0\\
    \,    &\text{sym}     &\,    &70.8 &0&0\\
   \,     &\,   &\,    &\,   &81.4&0\\
    \,    &\,    &\,    &\,    &\,    &74.5
    \end{bmatrix}.
\end{equation}

Unidirectional CFRP stiffness matrix in Voigt notation (units=GPa), $\rho= 1.6\ \mathrm{g/cm^3}$ \citep{nayfeh1995wave}.
\begin{equation}\label{eq:cfrp}
    \begin{bmatrix}
    155.43&3.72&3.72&0&0&0\\
    \,    &16.34&4.96&0&0&0\\
    \,    &\,    &16.34&0&0&0\\
    \,    &\text{sym}     &\, &3.37&0&0\\
    \,     &\,   &\,    &\,   &7.48&0\\
    \,    &\,    &\,    &\,    &\,    &7.48
    \end{bmatrix}.
\end{equation}
Single crystal silicon stiffness matrix in Voigt notation (units=GPa), $\rho= 2.23\  \mathrm{g/cm^3}$ \citep{Hall1967}.
\begin{equation}\label{eq:si}
    \begin{bmatrix}
    165.7&64&64&0&0&0\\
    \,    &165.7&64&0&0&0\\
    \,    &\,    &165.7&0&0&0\\
    \,    &\text{sym}     &\, &79.6&0&0\\
    \,     &\,   &\,    &\,   &79.68&0\\
    \,    &\,    &\,    &\,    &\,    &79.6
    \end{bmatrix}.
\end{equation}
\section{Material properties from PWM fitting}\label{appendix:2}
Stainless steel 304 grade stiffness matrix in Voigt notation (units=GPa) $\rho= 7.8\  \mathrm{g/cm^3}$
\begin{equation}\label{eq:ss_fit}
\begin{bmatrix}
    269.1&103.7&104.6 &0&0&0\\
    \,    &265.9&107.0 &0&0&0\\
    \,    &\,    &  261.9 &0&0&0\\
    \,    &\text{sym}     &\,    &78.1 &0&0\\
   \,     &\,   &\,    &\,   &757&0\\
    \,    &\,    &\,    &\,    &\,    &73.3
    \end{bmatrix}.
\end{equation}

Unidirectional CFRP stiffness matrix in Voigt notation (units=GPa) $\rho= 1.41\  \mathrm{g/cm^3}$
\begin{equation}\label{eq:cfrp_fit}
    \begin{bmatrix}
    233.1&2.6&2.7&0&0&0\\
    \,    &7.9&4.5&0&0&0\\
    \,    &\,    &7.6&0&0&0\\
    \,    &\text{sym}     &\, &1.6&0&0\\
    \,     &\,   &\,    &\,   &4.0&0\\
    \,    &\,    &\,    &\,    &\,    &4.3
    \end{bmatrix}.
\end{equation}

\bibliography{references}

\end{document}